\documentclass[sigconf]{acmart}
\usepackage{adjustbox}
\usepackage{multirow}
\setcopyright{none}
\renewcommand\footnotetextcopyrightpermission[1]{} 
\acmConference[MM '22]{Proceedings of the 30th ACM International Conference on Multimedia}{October 10--14, 2022}{Lisboa, Portugal}
\AtBeginDocument{%
  \providecommand\BibTeX{{%
    \normalfont B\kern-0.5em{\scshape i\kern-0.25em b}\kern-0.8em\TeX}}}

\begin{document}

\title{Multiview Contrastive Learning for Completely Blind Video Quality Assessment of User Generated Content}


\author{Shankhanil Mitra}
\affiliation{%
 \institution{Indian Institute of Science}
 \streetaddress{C.V.Raman Road}
 \city{Bangalore}
 \state{Karnataka}
 \country{India}}

\author{Rajiv Soundararajan}
\affiliation{%
 \institution{Indian Institute of Science}
 \streetaddress{C.V.Raman Road}
 \city{Bangalore}
 \state{Karnataka}
 \country{India}}

\renewcommand{\shortauthors}{Mitra, et al.}

\begin{abstract}
  Completely blind video quality assessment (VQA) refers to a class of quality assessment methods that do not use any reference videos, human opinion scores or training videos from the target database to learn a quality model. The design of this class of methods is particularly important since it can allow for superior generalization in performance across various datasets. We consider the design of completely blind VQA for user generated content. While several deep feature extraction methods have been considered in supervised and weakly supervised settings, such approaches have not been studied in the context of completely blind VQA. We bridge this gap by presenting a self-supervised multiview contrastive learning framework to learn spatio-temporal quality representations. In particular, we capture the common information between frame differences and frames by treating them as a pair of views and similarly obtain the shared representations between frame differences and optical flow. The resulting features are then compared with a corpus of pristine natural video patches to predict the quality of the distorted video. Detailed experiments on multiple camera captured VQA datasets reveal the superior performance of our method over other features when evaluated without training on human scores. Code will be made available at \url{https://github.com/Shankhanil006/VISION}.
\end{abstract}

\ccsdesc[500]{Computing methodologies~Computer vision}
\ccsdesc[500]{Computing methodologies~Image processing}
\keywords{Multiview contrastive learning, blind video quality assessment, user generated content}

\maketitle

\section{Introduction}
The ubiquity of mobile devices and video sharing platforms has led to an explosion in the number of videos captured, processed, and shared. Thus, the perceptual quality assessment of such videos is of paramount importance in enabling a better user experience. The video quality assessment (VQA) of such user generated content obtained through camera captures is challenging on multiple counts. The authentic distortions that arise in user generated content are more complex and often suffer from multiple sources of video degradation at the same time. Further, unavailability of reference video motivates the study of no reference (NR) VQA.

NR VQA has been studied quite extensively in the supervised setting \cite{videval, qa_in_the_wild, rirnet, tlvqm, cnntlvqm} for user generated content (UGC), where video features are regressed against human scores to learn a video quality model. However, such an approach suffers from two main drawbacks. Firstly, it requires a large number of human ratings to train a model. Secondly, the generalization performance of such supervised approaches has also been found to be a limitation \cite{mdtvsfa}. This motivates the need for completely blind NR VQA approaches where neither the human ratings nor the videos from the target database are used in any step of the algorithm design. 

The problem of designing completely blind NR VQA where neither human scores nor the videos in the target database are used has been researched to some extent. The VIIDEO \cite{viideo} model represents one such example where the models based on the statistics of natural videos are designed to measure video quality. Recently, the perceptual straightening hypothesis of the temporal information of natural videos has been used to design completely blind NR VQA \cite{STEM}. The natural image quality evaluator (NIQE) \cite{niqe} which is a completely blind image quality metric has also had limited success in NR VQA. Nevertheless, the role of deep features for completely blind NR VQA has not been explored to the best of our knowledge. 

In this work, we explore self-supervised contrastive learning of video quality features and deploy them to predict video quality without training on human labels. Self-supervised learning for video quality representations has been explored recently using predicted frames as augmentations for contrastive learning \cite{cspt}. However, the inaccuracy of the video prediction algorithm can limit the performance of the quality representation learning. Further, the learnt features have only been evaluated in a supervised setup as opposed to an unsupervised prediction of quality we explore in our work. HEKE \cite{heke-csvt} learns spatio-temporal representations from synthetically distorted videos suffering from artifacts due to compression and packet losses. But such a model may not perform well on authentic distortions which arise during camera captures. 

Our main contribution in this work is the design of a multiview contrastive learning framework for quality representation learning. We refer to our framework and the resulting quality metric as Video quality Index uSing multIview cOntrastive learNing \textbf{(VISION)}. We observe that frame differences are often interpreted as moving edges and contain information about the spatial and temporal quality. Thus, their joint distribution with frames and optical flow can be used to extract interesting quality related features. In particular, we capture the joint distribution of quality features in frames and frame differences for predicting video quality. Similarly, we also extract features that correspond to the joint distribution of frame differences and optical flow for quality prediction. The two sets of features are then compared against a corpus of such features from pristine videos to predict video quality. 

We conduct detailed experiments on multiple datasets to show the effective performance of our model. In particular, we also design novel benchmarks for comparison, where we evaluate several features learnt using a variety of approaches for unsupervised quality prediction without training them on human scores. This analysis is interesting in understanding the performance of the features for predicting video quality without human training. 

In summary, the main contributions of this work are as follows:

\begin{itemize}
    \item We design a multiview contrastive learning approach for learning quality representations for user generated content. In particular, our features capture the joint distributions of frame differences with frames and optical flow. 
    \item We show through detailed experiments on multiple datasets that the learnt features can be used to effectively predict video quality by comparing with a corpus of such features from pristine videos. 
    \item We evaluate the performance of several interesting features that can be learnt without human scores and evaluated for opinion unaware quality assessment by comparing against a corpus of pristine videos. 
\end{itemize}

\section{Related Work}

\textbf{Supervised NR VQA:} 
One of the most successful NR VQA approaches models natural scene statistics to generate quality aware features. Handcrafted methods involve designing a statistical model based upon discrete cosine transform (DCT) coefficients of frame differences \cite{vbliind}, 3D DCT coefficients \cite{3d_dct}, and 3D mean subtracted contrast normalized coefficients \cite{nstss} of video clips, and optical flow based \cite{optical_nrvqa}. Several other NR VQA methods have been developed by taking into account blockiness, sharpness, noise, and temporal correlation in videos \cite{nrvqa_1,nrvqa_2,nrvqa_3}. TLVQM \cite{tlvqm} deploys low complexity and high complexity features from video frames based on spatial, and temporal statistics of videos. VIDEVAL \cite{videval} combines handcrafted features from several existing blind VQA algorithms.

In recent years, convolutional neural networks (CNN) have been used to extract quality aware features from videos. One of the approaches involves combining CNN based features with other heuristics based features to achieve state-of-the-art NR VQA performance \cite{rapique, cnntlvqm,cnnvqaicip2018}. On the other hand, an end-to-end deep learning based method was designed to predict compressed video quality for specific codecs \cite{vmeon}. Motion representation based models \cite{rirnet}  have also been studied for NR VQA. Features extracted by learning a 3D CNN model on video clips \cite{deep3dcnnvqa}, and pretrained ResNet50 \cite{resnet} features learnt on ImageNet database have been fed to a recurrent model to predict quality \cite{qa_in_the_wild}. Quality aware features have also been extracted from image quality models such a PaQ-2-PiQ \cite{paq-2-piq} and combined with pretrained 3D Resnet18 features to predict global video quality \cite{patchVQ}.  UCDA \cite{ucda} adopts a domain adaptation approach to adapt a model learnt on synthetic video distortions to authentic video distortions. While UCDA does not use any human quality scores for the target database, it trains on a large set of human annotated labels for the source database. 

\noindent\textbf{Pseudo-label Training for NR VQA:} A broad class of opinion unaware quality models is designed by learning to predict full reference metrics available in the training data. These models are relevant for synthetic distortions due to the need for a reference for generating the target quality index. Video CORNIA \cite{vcornia} 
was designed by first learning quality aware features through visual codebooks in an unsupervised manner and then regressing these features against frame level full reference quality measures. A weakly supervised approach was adopted for pretraining, where a deep network was first trained to predict a full reference quality measure. The features were then fine-tuned on human opinion scores \cite{weakly_sup}. Pseudo-labels from multiple full reference video quality measures are used to obtain a richer quality prediction method on account of the heterogeneity of the measures \cite{heke-csvt}. A similar approach is adopted based on the spatio-temporal entropic differences index exploiting the complementarity of the spatial and temporal streams \cite{nrsted}. 

\noindent\textbf{Completely Blind NR VQA:} The class of completely blind VQA algorithms is very challenging and very few algorithms have been designed. One of the first such methods was the VIIDEO method \cite{viideo}, which models the intrinsic statistical regularities in natural videos to measure the disturbances in the presence of distortions. STEM \cite{STEM} is a recent completely blind VQA model designed for user generated content that exploits the human perception of straighter temporal trajectories for natural videos in a transformed space. The loss in straightness or the resulting curvature in the presence of distortions is used to measure  quality. This approach is used in conjunction with the completely blind natural image quality evaluator to obtain impressive performance. 

\noindent\textbf{Self-supervised learning for image and video quality:} The role of self-supervised feature learning in VQA is still nascent. Contrastive self-supervised pretraining \cite{cspt} has been explored for learning deep video quality features for supervised NR VQA. In particular, frame prediction algorithms are deployed to create augmented videos which are used to learn quality distortion and content features for NR VQA. CONTRIQUE \cite{contrique} uses information about distortion levels and types of synthetically distorted images, along with authentically distorted images for supervised NR image quality assessment. Nevertheless, self-supervised methods are only being deployed in a supervised setup, while such approaches offer the possibility of completely blind VQA, which we explore in our work. 

\begin{figure*}
    \centering
    \includegraphics[width = \textwidth]{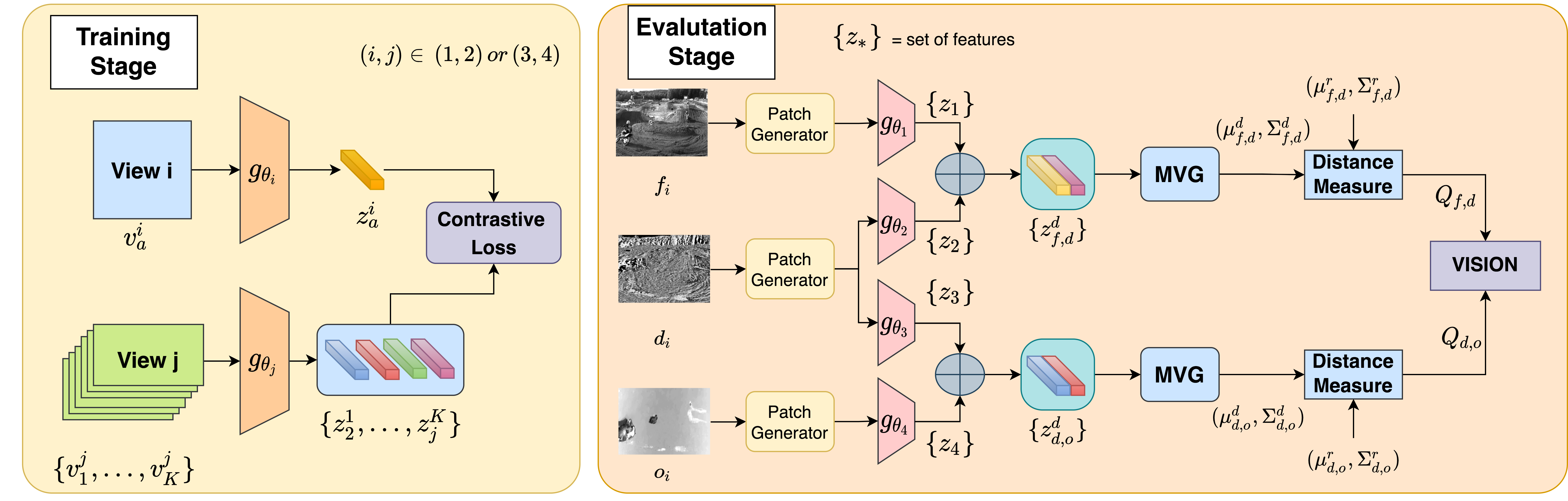}
    \caption{Illustration of VISION algorithm. Training Stage: For every sample of the first view, one positive and $(K-1)$ negative sample are chosen from the second view. Encoders $g_{\theta_i}$, and $g_{\theta_j}$ are optimized with contrastive loss using the positive and negative set of features. Evaluation Stage: Feature embeddings are extracted from frames, frame differences, and optical flow using the learned encoder in the training stage. Quality is predicted by measuring the distance in Equation (\ref{distance}) using the MVG model parameters of pristine and test video patches.}
    \label{fig:flowchart}
\end{figure*}

\section{Method}

\subsection{Overview}
 An overview of our quality aware representation learning from videos using multiple views of video frames without any human subjective score is shown in Figure \ref{fig:flowchart}. Our approach is inspired by the success of contrastive multiview coding (CMC) \cite{cmc} in image classification, where the goal is to elicit features that maximize the shared information contained in both the views. This can be achieved by maximizing the mutual information between the resulting features. However, we adapt the CMC method to create positive and negative view pairs that vary only in quality to learn rich quality representations through contrastive learning. The key question that arises in CMC is the choice of views. We note that frame differences in videos contain information about variations in quality across frames and thus some information about the spatial quality of the frames. Further, frame differences also contain temporal information and the common information with optical flow may be used to capture temporal quality. We capture the common information between frame differences and frames and similarly between frame differences and optical flow to predict video quality. We note that the choice of views in our framework for quality representation learning is different from the choice of views in the original CMC work \cite{cmc}. 
 In the following sections, we first discuss how we generate a set of videos, consisting of variations of the video that differ only in quality. We then discuss our multiview contrastive learning framework on this dataset. 
 

\subsection{Data Generation}
Two broad sets of videos are employed for learning our quality representations. We first consider databases of videos where pristine videos are corrupted with synthetic distortions such as compression,  and transmission errors. 
We employ existing synthetically distorted datasets like the LIVE Mobile \cite{mobile1}, LIVE VQA \cite{live_sd1}, EPFL-PoliMI \cite{epfl1}, CSIQ VQD \cite{csiq}, and ECVQ-EVVQ \cite{ecvq_evvq1} which contain pristine videos corrupted with rich synthetic distortions like MPEG-2, MPEG-4, H.264, noise, gaussian blur, IP loss, wavelet compression based snow codec, wireless transmission loss, and so on. Although our focus is on authentic distortions, we still believe that the synthetic distortions help learn good quality representations. Since our study mainly focuses on quality prediction for authentically distorted videos, we also generate different synthetically distorted versions of a camera captured video. 

We hypothesize that by creating variations of the authentically distorted video in terms of further quality degradations, and then learning features that contrast these videos, we can learn quality representations of the underlying authentically distorted video. In particular, we randomly sample a camera captured video from an authentically distorted video database namely LIVE Large-Scale Social Video Quality \cite{patchVQ}. We generate augmentations of this video with MPEG-2, and H.264 compression at different quality levels, and downsampling and upsampling at different scales. In addition, we use frame interpolation techniques \cite{ffmpeg} to create augmentations by first reducing the frame rate of the video and then generating the video at the original frame rate through interpolation. We create a combined database of synthetically distorted videos with compression and transmission errors, and also a database of authentically distorted videos with its various distorted versions. 

\subsection{Multiview Contrastive Feature Learning}
As we remarked earlier, we choose two pairs of views, one between frame differences and frames and another between frame differences and optical flow to learn our quality representations. We note that there does not appear to exist any shared information between frames and optical flow and do not consider that pair. Thus, we constitute a two stream approach to learn quality aware representation. In the first stream, frame $f$ and corresponding frame difference $d$ with respect to its neighbour is used to form the two views of a multiview contrastive framework. In the other stream, frame difference $d$ and optical flow $o$ are considered as two views of a CMC \cite{cmc} framework. Our quality representation learning framework is temporally localized. 


We first describe the multiview representation learning with frames and frame differences as the pair of views. The multiview contrastive learning framework requires a pair of congruent and several incongruent pairs to learn quality representations. While the congruent pairs are chosen as the feature representations from frames and frame differences of the same video, the incongruent pairs are chosen as frames and frame differences from different videos having the same content but different distortions. This is a key aspect of our model where the positives and negatives are always drawn from the same scene.  In a mini-batch, we sample $S$ scenes from the training database, and for each scene, we have a set of $K$ videos with different distortions. Let  $\{ V_1^s, V_2^s, \ldots, V_K^s \}$ be a set of $K$ videos with different distortion types and levels but having the same content, where, $s = \{ 1, 2, \ldots, S \}$. Let, $f_j$ and $d_j$ denote the frame and  frame difference (with frame $f_j$ as one of the frames)  of the video $V_j^s$ respectively at a certain time instance where $j\in\{1,2,\ldots,K\}$. We describe the training method for $V_j^s$ below and the same can be extended to other videos. Let frame $f_j$ be chosen as the anchor view. The positive or congruent pair of views corresponding to this anchor is $\{f_j, d_j\}$, while the negative pairs are given by $\{f_j, d_k\}_{k \ne j}$ by enumerating over the frame differences of all incongruent videos.

We deploy two CNNs, $g_{\theta_1}(.)$ and $g_{\theta_2}(.)$ with the same architecture but different parameters $\theta_1$ and $\theta_2$ to learn feature representations of the frames and frame differences, respectively. A discriminative function $h_\theta(.)$ is trained to give high similarity between $g_{\theta_1}(f_j)$, and $g_{\theta_2}(d_j)$ and low similarity between $g_{\theta_1}(f_j)$ and $g_{\theta_2}(d_k)$, where $k\neq j$ and $k\in\{1,2,\ldots,K\}$. 
Let, the feature representation for frame $f$ and frame difference $d$ be given by,
\begin{equation*}
    z_1 = g_{\theta_1}(f) \  ,  \  z_2 = g_{\theta_2}(d).
\end{equation*}
The cosine similarity between the embedding $z_1$ and $z_2$ is given by:

\begin{equation}
h_\theta(f, d) = \exp\left(\frac{z_1^T z_2}{|| z_1|| ||z_2||}. \frac{1}{\tau}\right),
\end{equation}
where $\tau$ is the dynamic range adjuster.
Therefore, the contrastive loss can be written as:

\begin{equation}
    l(f_j,d_j) = -\log \frac{h_\theta(f_j, d_j)}{\sum_{k=1}^K h_\theta(f_j, d_k)}
\end{equation}
Enumerating the anchor view over all the videos in $\{ V_1^s, V_2^s, ..., V_K^s \}_{s=1}^S$, we have the overall loss term with frames taken as anchor view as,
\begin{equation}
    l(f,d) = \frac{1}{SK}\sum_{s=1}^S \sum_{j=1}^K l(f_j, d_j)
    \label{loss_view}
\end{equation}

Similarly, taking the anchor view as frame differences $d_j$, we have,
$l(d,f) = \frac{1}{SK}\sum_{s=1}^S \sum_{j=1}^K l(d_j, f_j)$. We obtain the overall two view loss function as
\begin{equation}
    \mathcal{L}_{f,d} = l(f,d) + l(d,f). 
    \label{loss_fd}
\end{equation}

For the second stream in our contrastive learning setup, we choose frame difference $d$ and optical flow map $o$ at a certain time instance of a video as the two views. Similar to the above approach, we choose two CNN based encoder networks $g_{\theta_3}(.)$ and $g_{\theta_4}(.)$ to learn the representations of $d$ and $o$, respectively. Let $z_3$, and $z_4$ represent the feature output of $g_{\theta_3}(.)$ and $g_{\theta_4}(.)$. Then the similarity between these embeddings is given as,
\begin{equation}
    h_\theta(d, o) = \exp\left(\frac{z_3^T z_4}{|| z_3|| ||z_4||} . \frac{1}{\tau}\right).
\end{equation}
Taking frame differences as the anchor view and enumerating them over optical flow for positive and negative pairs like Equation (\ref{loss_view}), we have the contrastive loss as
\begin{align}
    l(d_j,o_j) = - \log \frac{h_\theta(d_j, o_j)}{\sum_{k=1}^K h_\theta(d_j, o_k)}\\
    l(d,o) = \frac{1}{SK}\sum_{s=1}^S \sum_{j=1}^K l(d_j, o_j).
\end{align}
The overall objective function with either of the views taken as anchor view is given as,
\begin{equation} 
    \mathcal{L}_{d,o} = l(d,o) + l(o,d). 
    \label{loss_do}
\end{equation}

We train all the four networks in the two streams with the respective loss functions in Equation (\ref{loss_fd}), and (\ref{loss_do}). Ideally, either of the CNNs from a given pair of views may be used to extract features during the prediction stage. However, we note that using both CNNs may be beneficial in overcoming any residual errors in perfectly contrasting the positive and negative pairs. Thus,   during the prediction stage, we obtain the embedding for the first stream with frame and frame difference as views as $z_{f,d} = (z_1 + z_2)/2$, and for the second stream as $z_{d,o} = (z_3 + z_4)/2$. In Section \ref{ssec:averaging}, we provide a detailed study of using the average of the feature representation over using individual features.

\subsection{Distance Measure and Quality Prediction} \label{ssec:niqe}
The goal of our work is to estimate the quality of videos in a completely blind setup. We choose a distance based approach similar to NIQE \cite{niqe} to predict quality given the feature representation of a distorted video and a corpus of pristine videos. Since we extract two sets of features $z_{f,d}$ and $z_{d,o}$ from the video, we compute the distance with respect to each of these features and combine them. We first describe the distance computation using $z_{f,d}$. A similar approach is adopted for $z_{d,o}$. 

We generate patches of size $R\times R$ from the frames of pristine videos corresponding to each of the views, i.e. frame and frame difference. Similar to NIQE \cite{niqe}, we select those patches in each frame that have a sharpness greater than $\tau_s$ times the sharpness of the sharpest patch of the frame. The frame differences are also drawn at the corresponding same locations as the sharp patches. A multivariate Gaussian (MVG) model with parameters $(\mu_r^{fd}, \Sigma_r^{fd})$ is learnt on the feature representation $z_{f,d}$ of these sharp patches drawn from the set of pristine videos. We now predict the quality of the $i^{th}$ frame in a distorted video as follows. We extract $R \times R$ patches with no overlap from $i^{th}$ frame and obtain the feature embeddings for all the patches similar to NIQE \cite{niqe}. An MVG model with parameters $(\mu_d^{fd}, \Sigma_d^{fd})$ is learnt on these distorted patches. The quality estimate of the $i^{th}$ frame in a distorted video is given as,
\begin{equation}
    q^i_{f,d} = \sqrt{(\mu_r^{fd} -\mu_d^{fd})^T \left(\frac{\Sigma_r^{fd} +\Sigma_d^{fd}}{2} \right)^{-1} (\mu_r^{fd} - \mu_d^{fd})}. 
    \label{distance}
\end{equation}

We compute the above quality index for frames sampled at 1 frame per second in the distorted video. For a video of duration $N$ seconds, $i\in\{1,2,\ldots,N\}$, the video level quality is estimated by average pooling the frame level quality predictions as,
\begin{equation}
    Q_{f,d} = \frac{1}{N}\sum_{i=1}^N q^i_{f,d}.
    \label{vid2frame_qa}
\end{equation}
Similarly, we extract the embedding $z_{d,o}$ using the second pair of networks corresponding to frame differences and optical flow maps from the same locations of sharp patches determined above. Let $q^i_{d,o}$ denote the quality of a video at the $i^{th}$ time instance using $z_{d,o}$. The video level quality from the second stream is given by,
\begin{equation}
    Q_{d,o} = \frac{1}{N}\sum_{i=1}^N q^i_{d,o}. 
\end{equation}

\textbf{VISION}. The overall video quality for the test video is given as the product of the predicted quality using the two streams as, 
\begin{equation}
    VISION = Q_{f,d}*Q_{d,o}.
    \label{vision}
\end{equation}
Our overall quality prediction using the product of individual components is similar to that of ST-RRED \cite{strred}, and NR-STED \cite{nrsted}.

\section{Experiments and Results}

\subsection{Experimental Settings}

\textbf{Training Data}. \label{ssec:data} We train our model in a self-supervised fashion on both synthetically and authentically distorted datasets.

\textbf{\textit{Synthetic Database}}. We use a combination of 5 databases comprising 850 distorted videos produced from 60 pristine videos at different resolutions and frame rates. The details of distortions present in the synthetic databases are as follows:

\textbf{LIVE VQA} \cite{live_sd1}: The videos suffer from compression artifacts due to MPEG-2 and H.264 as well as transmission distortions obtained by sending H.264 compressed videos through error-prone IP and wireless networks. 

\textbf{ LIVE Mobile} \cite{mobile1}: A set of videos comprising compression, wireless packet-loss, rate-adaptation, and temporal dynamics are included in this database. 

\textbf{ EPFL-PoliMI} \cite{epfl1}: This database is composed of videos at two different resolutions encoded with H.264/AVC and passed through an error prone channel. 

\textbf{ CSIQ VQD} \cite{csiq}: Here, we have videos comprising H.264, HEVC, Motion JPEG, wavelet based compression using Snow codec and H.264 were subjected to wireless transmission channel. 

\textbf{ ECVQ and EVVQ} \cite{ecvq_evvq1}: The artifacts include H.264 and MPEG-4 visual compression.
    
\textbf{\textit{Authentic Database}}. Since our objective in this work is completely blind estimation of video quality of user generated content, which is typically authentically distorted, we use videos from the LIVE-FB Large-Scale Social Video Quality (LSVQ) \cite{patchVQ} database for our training. Since our model works with frame level data, we randomly sample 200 videos from the 39K videos in the LSVQ video database.  LSVQ contains only a single distorted video for each content, so we augment the video with further distortions for contrastive learning. For distortion augmentation, we use MPEG-2, H.264, video downsampling followed by upsampling, and video frame interpolation. For every distortion type, we generate videos at three different distortion levels. We use ffmpeg \cite{ffmpeg} to generate MPEG-2 distorted videos with \emph{qscale} index varying between 1 to 20. Similarly, to corrupt the authentic videos with H.264, we use ffmpeg \cite{ffmpeg} at a \emph{crf} value between 10 to 50. Sampling is done by downscaling the video at a rate of $2, 4$, and $8$ and upscaling them back to original resolution. For frame interpolation, we read each video at $0.25, 0.33$, and $0.5$ times the original frame rate. Then we fill the frame by interpolation using \emph{minterpolate} filter in ffmpeg \cite{ffmpeg} to get the distorted video at the original frame rate.

\textbf{Details of Views}
The input data to the encoder $g_{\theta_1}(.)$ consists of grayscale frames, while the input to the encoders $g_{\theta_2}(.)$, and $g_{\theta_3}(.)$ consists of the difference of grayscale frames. The input to $g_{\theta_4}(.)$ is a two channel optical flow map, each channel representing horizontal and vertical displacement.

\textbf{Training Details}
The encoder chosen to extract features from frames, frame differences, and optical flow viz. $g_{\theta_1}(.)$, $g_{\theta_2}(.)$, $g_{\theta_3}(.)$, and $g_{\theta_4}(.)$ have the same architecture and each of them outputs a $256$ dimensional feature vector. The encoder architecture is given in the supplementary material. Each of the encoders takes a frame level input at a certain time instance. The optical flow map is estimated using TV-L1 algorithm \cite{tvl1}. The training was done using a batch size $S = 8$ with Adam optimizer \cite{adam} at a learning rate of $1e-4$ for 5000 iterations. For each batch of input, we have $1$ positive and $10$ negative pairs. Due to the computational complexity, we centre crop the views, taking the input at a resolution of $224 \times 224$. The dynamic range adjuster is chosen to be 0.1 similar to \cite{contrique, simclr}. 

\textbf{Evaluation Databases} Since our goal is to design a completely blind VQA algorithm for user generated content, we omit the LSVQ \cite{patchVQ} dataset for quality prediction of authentically distorted videos since we trained on that dataset. We conduct experiments on four user generated datasets as follows:

\textbf{ KoNViD-1K} \cite{konvid}: This dataset contains 1200 videos
filtered from the YFCC100m database consisting of 793436
sequences. The videos in this database contain a wide variety
of content, distortion types, and subjective quality variations.
The videos are of $720 \times 540$ resolution, corresponing to a frame rate of 24, 25, or 30 frames per second, and 8 seconds in duration.

\textbf{LIVE Video Quality Challenge (LVQC) Database}\cite{livevqc}:
The LIVE VQC database consists of 585 videos of unique
content captured from 101 different devices leading to a
widespread of complex authentic distortions. LVQC has 10 second long videos available at 18 different spatial resolutions ranging between $1980 \times 1080$ to $320 \times 240$ across landscape and portrait modes. 

\textbf{LIVE Qualcomm Database (LQCOMM)}\cite{liveqcomm}: This database consists of 208 videos accounting for distortions generated during the camera capture process using eight mobile devices. The videos are of spatial resolution $1920 \times 1080$, 15 seconds long when played at 30 fps.

\textbf{YouTube-UGC} \cite{youtube_ugc}: This database contains 1380 user generated videos at resolutions varying between 360p to 4k. The videos are 20 seconds long in duration. This database contains videos belonging to 15 categories (e.g. gaming, sports, and music videos).

\begin{table*}
\caption{Performance evaluation of VISION against other completely blind benchmarking algorithms on four user generated content datasets. Methods marked with $(*)$ are modified to predict quality in a blind fashion using the distance metric in Section \ref{distance}. The \textbf{\textit{Emphasised}}, and \textbf{Boldfaced} entries indicate the best and second-best performance in each database.}
\centering
\label{tab:main_table} 
\begin{tabular}{|c|c c|c c|c c|c c|}
\hline
& \multicolumn{2}{c|}{KoNVid-1K} & \multicolumn{2}{c|}{LIVE VQC} & \multicolumn{2}{c|}{Youtube-UGC}& \multicolumn{2}{c|}{LIVE Qualcomm} \\
\hline
Methods & SROCC & PLCC & SROCC & PLCC & SROCC & PLCC & SROCC & PLCC\\
\hline
VIIDEO  &0.013 & -0.015 & 0.029 & 0.137 & 0.130 & 0.146 & -0.141 & 0.098\\
$VCORNIA^*$ (1fr/sec) & 0.112 & 0.132  & 0.166 & 0.133 & 0.461 &0.455 & 0.186 & 0.267\\
NIQE (1 fr/sec) & 0.542 & 0.544 & 0.563 & 0.610 & 0.236 & 0.105 & 0.467 & 0.504\\
$ResNet50^*$ (1 fr/sec) & 0.273 & 0.288 & 0.240 & 0.275  & 0.466 & 0.465 & 0.313 & 0.362 \\
STEM & \textit{\textbf{0.629}} & \textit{\textbf{0.629}} & \textbf{0.656} & \textbf{0.670} & 0.284 & 0.318 & \textbf{0.483} & \textbf{0.537}\\
$HEKE^*$ (1 fr/sec) & 0.487 & 0.508 & 0.444 & 0.525 & \textbf{0.462} & \textbf{0.501} & 0.236 & 0.327\\
\hline
$Q_{f,d}$ (1 fr/sec) &0.545 & 0.558 & 0.549 & 0.592 & 0.492 & 0.501 & 0.441 & 0.479\\
$Q_{d,o}$ (1 fr/sec) &0.496 & 0.497 & 0.647 & 0.664 & 0.466 & 0.479 & 0.502 & 0.534\\
$VISION$ (1 fr/sec) & \textbf{0.598} & \textbf{0.597} & \textit{\textbf{0.676}} & \textit{\textbf{0.701}} & \textit{\textbf{0.503}} & \textit{\textbf{0.510}} & \textit{\textbf{0.547}} & \textit{\textbf{0.576}}\\
\hline
\end{tabular}
\end{table*}

\textbf{Evaluation Details}.
Similar to NIQE \cite{niqe}, we choose patches of size $96 \times 96$ ($R =96$). To generate pristine patches, we choose reference videos from \cite{live_sd1,mobile1,csiq,epfl1,ecvq_evvq1}. We choose a sharpness threshold $\tau_s = 0.85$ for generating the pristine set of patches for all our experiments and also for the other benchmarking algorithms. To reduce the computational capacity in estimating the optical flow map using the TV-L1 algorithm at test time, we estimate the optical flow at $1/8$th of the spatial resolution of the video and upsample the flow to the original resolution with appropriate scaling. Since the computational time required to predict video quality is large if the resolution is greater than 720p as shown in prior work \cite{videval, rapique, tlvqm}, we see that quality prediction at 1 frame per second followed by averaging pooling of the scores gives a similar performance to taking all the frames for prediction. Thus, we compute the quality of videos at 1 frame per second for VISION.

We evaluate the performance of the completely blind VQA methods using the conventional measures such as Spearman’s rank order correlation coefficient (SROCC), and linear correlation coefficient (LCC) between the predicted quality scores and the ground truth quality scores. Since no training is required on the test dataset, we do not need any training or validation splits of the test dataset. All the videos in the test datasets were used for evaluation. We pass our predicted quality through a four parameter monotonic logistic function given by, $Q_{trans} = \beta_2 + \frac{\beta_1 - \beta_2}{1 + \exp (-((Q-\beta_2)/\beta_4)}$ as in \cite{nstss, STEM, nrsted}, where $Q$ is the predicted quality before computing the performance measures.

\subsection{Benchmarking Algorithms}

We compare VISION with other completely blind VQA algorithms like VIIDEO \cite{viideo} and STEM \cite{STEM}. Since our quality estimation using the distance metric in Equation (\ref{distance}) is applied at a frame level, we extend our benchmarking analysis to completely blind IQA methods like NIQE \cite{niqe} to compute frame level quality. We estimate the video quality using NIQE \cite{niqe} as in Equation (\ref{vid2frame_qa}) at 1 frame/sec. In addition to the above, we present several interesting benchmarks where we evaluate methods that learn features without human supervision and use them to evaluate quality by comparing with a corpus of pristine videos. We note that VCORNIA \cite{vcornia} learns frame level quality features without supervision. So, we apply our distance measurement criteria to the set of pristine and test video features generated using VCORNIA \cite{vcornia}. HEKE \cite{heke-csvt} learns quality aware features using synthetically distorted databases by regressing against pseudo-quality scores generated by various full reference metrics. Since the training of HEKE \cite{heke-csvt} is blind with respect to the authentic data, we also extract pristine and test video features using this model. Thereafter, we apply our distance measure to estimate the quality at a 1 frame per second and average pool the scores like our method. 
Pretrained ResNet50 \cite{resnet} features extracted at frame level perform well for NR VQA \cite{videval}. In our experiment, we extract ResNet50 features at patch level like VISION from the pristine corpus and the test video to measure quality of the test video using our distance measure. We term VCORNIA \cite{vcornia}, HEKE \cite{heke-csvt}, and pretrained ResNet50 \cite{resnet} formulated in a blind setup as $VCORNIA^*$, $HEKE^*$, and $ResNet50^*$ respectively.

\subsection{Performance Analysis}

From Table \ref{tab:main_table}, we see that VIIDEO \cite{viideo} performs very poorly on the authentically distorted videos across all databases. While $HEKE^*$ (1 fr/sec), and NIQE (1fr/sec) \cite{niqe} perform well on KoNVid-1K and LIVE VQC, NIQE (1fr/sec) performs poorly on Youtube-UGC \cite{youtube_ugc}, and $HEKE^*$ (1fr/sec) performs poorly on LIVE Qualcomm \cite{liveqcomm}. Though $VCORNIA^*$ (1fr/sec), and pretrained $ResNet50^*$ (1 fr/sec) performs well on the Youtube-UGC \cite{youtube_ugc} dataset, its performance on other datasets is poor.
We see that VISION gives comparable performance against the current state-of-the-art completely blind algorithm STEM \cite{STEM} on KoNVid-1K \cite{konvid}, and LIVE VQC \cite{livevqc} databases. On YouTube-UGC \cite{youtube_ugc}, and LIVE Qualcomm \cite{liveqcomm}, VISION outperforms STEM \cite{STEM}. Overall, VISION achieves a good performance consistently across all datasets. 

\subsection{Ablation Studies}

\begin{table}
\caption{Ablation analysis 
on different choice of views measured through SROCC. In the first column "View" denotes the views considered for learning in the multiview setup.}
\centering
\label{tab:averaging}
\begin{tabular}{|c|c|c|c|c|c|}
\hline
Views & Feature & KoNVid-1K & LVQC  & LQCOMM \\ 
\hline
Frame - & $z_1$ & 0.498 & 0.485  & 0.434 \\ 
Frame & $z_2$ & 0.507 & 0.473  & 0.382 \\ 

Diff. & $z_{f,d}$ & \textbf{0.545} & \textbf{0.549}  & \textbf{0.441} \\
\hline
\hline
Optical Fl. & $z_3$ & 0.371 & 0.568  & 0.480 \\

- Frame & $z_4$ & 0.472 & 0.575  & 0.372 \\
Diff. & $z_{d,o}$ & \textbf{0.496} & \textbf{0.647}  & \textbf{0.502} \\
\hline
VISION & $(z_{f,d} + z_{d,o})/2$ & 0.598 & 0.676 & 0.547\\ \hline
CMC & - & 0.401 & 0.492 & 0.443 \\
\hline
\end{tabular}
\end{table}

\textbf{Impact of Averaging the features from Multiple Views}. \label{ssec:averaging}
In Table \ref{tab:averaging}, we analyze whether the average of the features extracted from $g_{\theta_1}(.)$, and $g_{\theta_2}(.)$ gives better performance than the individual features extracted. A similar experiment is also conducted for $g_{\theta_3}(.)$, and $g_{\theta_4}(.)$.  Averaging tends to often give a better performance than the individual performance as seen in Table \ref{tab:averaging}. Thus we choose the averaged features from each stream to get $Q_{f,d}$, and $Q_{d,o}$. We also learned representations based on the views in CMC \cite{cmc} and presented the results in Table \ref{tab:averaging}. We infer that the views with our approach perform significantly better in learning  representations for the blind VQA task.

\textbf{Performance on Synthetically Distorted Databases}. We evaluate the performance of VISION on synthetically distorted databases such as LIVE VQA \cite{live_sd1}, LIVE Mobile \cite{mobile1}, CSIQ VQD \cite{csiq}, and EPFL-PoliMI \cite{epfl1}. Since, VISION was trained using all these synthetic databases, for this analysis, we leave out the test database during the training and the computation of the model for the pristine corpus.
Since there are very few unique scenes in the synthetic datasets and many videos are distorted versions of these, we find that all the methods are sensitive to the content causing a drop in performance. Nevertheless, VISION gives a more stable performance than other benchmarking algorithms on most of the databases.

\begin{table}
\caption{SROCC performance analysis on four synthetically distorted video databases. $VCORNIA^*$, NIQE \cite{niqe}, $ResNet50^*$,  $HEKE^*$, and VISION are applied at 1 frame/second for quality prediction.}
\centering
\label{tab:synthetic}
\begin{tabular}{|c|c|c|c|c|}
\hline
Methods  & CSIQ VQD & EPFL-Po & LIVE VQA & LIVE Mobile \\
\hline
VIIDEO &0.02 & 0.205 & \textbf{0.624} & 0.216\\
$VCORNIA^*$ & -0.02 & 0.05 & 0.15 & 0.289 \\
NIQE  & 0.440 & 0.185 & 0.174 & 0.427 \\
$ResNet50^*$ & 0.164 & 0.005 & 0.070 & 0.176\\
STEM &0.380 & 0.199 & 0.205 & 0.361\\
$HEKE^*$ & 0.291 & \textbf{0.347} & -0.139 & 0.370 \\
VISION & \textbf{0.463} & 0.225 & 0.273 & \textbf{0.433}\\

\hline
\end{tabular}
\end{table}

\textbf{Learning on Synthetic vs Authentic Datasets}.
VISION is learnt on a mix of synthetically distorted videos \cite{live_sd1,mobile1,epfl1,csiq,ecvq_evvq1}, and authentically distorted videos generated from LSVQ \cite{patchVQ}. We study the impact of learning with synthetic distortions and authentic distortions individually. In this experiment, VISION is trained with either synthetically distorted or authentically distorted videos. We show a comparative study of our model on three UGC test databases \cite{konvid,livevqc,liveqcomm} in Table \ref{tab:syn vs auth}. Though, learning on synthetic or authentic distortions alone gives a similar performance, the combined learning on mixed data gives superior performance on most databases. Combining synthetically distorted videos with the authentically distorted videos benefits the performance since the synthetically distorted databases contain richer sets of distortions generated by various study groups as mentioned in Section \ref{ssec:data}. 

\begin{table}
\caption{SROCC performance analysis of VISION on three UGC datasets when trained on synthetically distorted, authentically distorted and combined database.}
\centering
\label{tab:syn vs auth}
\begin{tabular}{|c|c|c|c|c|}
\hline
Data Type & KoNVid-1K & LIVE VQC & LIVE QCOMM\\
\hline

\multirow{1}{*}{Synthetic} 
& 0.592 & 0.630 & 0.524 \\
\hline 
\multirow{1}{*}{Authentic} 
& 0.567 & 0.667 & 0.542 \\
\hline 
\multirow{1}{*}{Combined} 
& \textbf{0.598}  & \textbf{0.676}  & \textbf{0.547} \\
\hline
\end{tabular}
\end{table}

\begin{figure}
    \centering
    \includegraphics[width=\columnwidth]{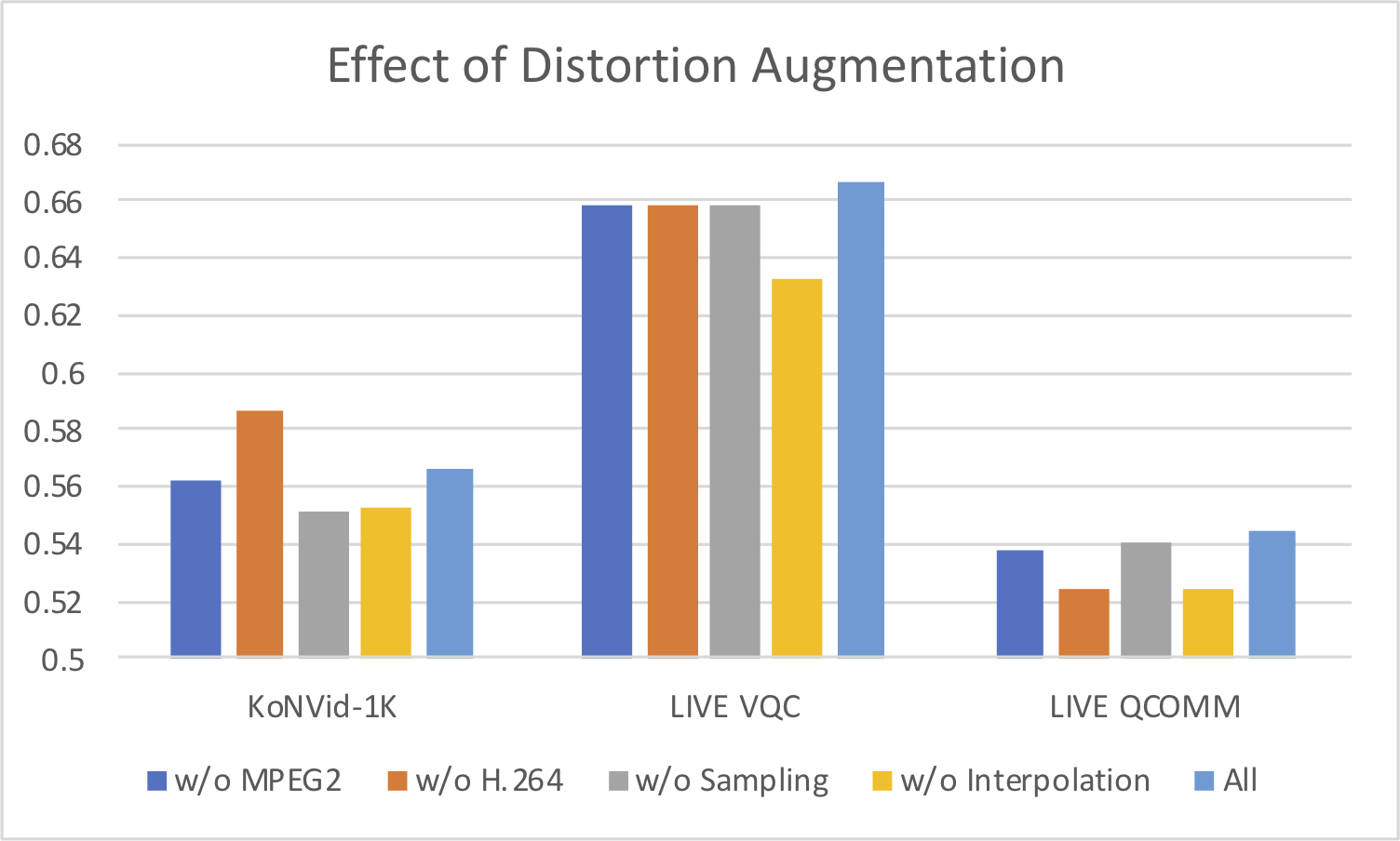}
    \caption{Effect of distortion augmentation on VISION performance. First four bars denote the performance measured by SROCC without one of the four distortion types mentioned in Section \ref{ssec:data}, while the last bar denotes the performance when all distortions are considered during training.}
    \label{fig:augmentation}
\end{figure}

\textbf{Impact of Distortion Augmentation}. To generate distorted views of the authentically distorted videos, we corrupt each UGC video with synthetic distortions such as MPEG-2, H.264, sampling, and interpolations as described in Section \ref{ssec:data}. We study the impact of synthetic augmentations when our model is trained on LSVQ \cite{patchVQ} videos corrupted with synthetic distortions, without one of the above four distortion types. In Figure \ref{fig:augmentation}, a comparative study on the performance of VISION is given when the encoders are trained with videos corrupted with all the above four distortion types vs when trained on videos corrupted with all but one distortion type. Overall, we see that no single augmentation is crucial and we get roughly similar performances even if we remove one of the augmentations.  

\begin{table}
\caption{SROCC performance comparison with other self-supervised video representation learning methods on three UGC datasets.  Entry marked '-' denotes data is unavailable.}
\centering
\label{tab:lin_eval_ssl}
\begin{tabular}{|c|c|c|c|c|}
\hline
Data Type & KoNVid-1K & LIVE VQC & LIVE QCOMM\\
\hline
DPC  & 0.426 & 0.485 & 0.308\\
VCOP  & 0.452 & 0.494 & -\\
PRP  & 0.349 & 0.415 & 0.281\\
IIC  & 0.527 & 0.534 & 0.361\\
CSPT  &0.702 & 0.623 & -\\
VISION & \textbf{0.724} & \textbf{0.698} & \textbf{0.623}\\
\hline 
\end{tabular}
\end{table}

\textbf{Self-Supervision based Video Representation Learning}. We also provide a comparative study of various self-supervision based video representation learning methods on three UGC databases in Table \ref{tab:lin_eval_ssl}. DPC \cite{dpc_ssl}, VCOP \cite{vcop_ssl}, PRP \cite{prp_ssl}, and IIC \cite{iic_ssl} learn spatio-temporal video features by learning an action recognition task. CSPT \cite{cspt} is a contrastive learning based VQA method modelled on learning quality based features by pretraining on a future video frame prediction task. 
We train a linear regressor (ridge regression) on top of the video level features extracted from all the above mentioned self-supervised algorithms and VISION (average pooled across frames) except CSPT \cite{cspt}, and VCOP \cite{vcop_ssl}. 
We conduct this test on 100 splits by dividing each of the 3 UGC datasets in the ratio of $80:20$ for the train-test splits. The numbers reported for CSPT \cite{cspt}, and VCOP \cite{vcop_ssl} are taken from literature. We see that VISION outperforms these self-supervised algorithms in a linear evaluation testing protocol. Thus, our multiview contrastive learning based approach using sets of distorted video samples with similar content learns good features for quality prediction.  

\textbf{Linear Evaluation of VQA Methods}.
We benchmark the quality aware feature representation following the linear evaluation protocol for SOTA feature based VQA methods. We divide three different authentically distorted datasets, each in the ratio of $80:20$ for the train-test splits and train a linear regressor 
for 100 splits. In Table \ref{tab:lin_eval_vqa}, we see VISION gives a competitive performance with other benchmarking methods on KoNVid-1K \cite{konvid}, and LIVE VQC \cite{livevqc}, while on LIVE Qualcomm \cite{liveqcomm} it is slightly lower than some other methods. We note that although the pretrained ResNet50 model achieves better performance than VISION in a linear supervised setting, in a completely blind setup, VISION features completely outperform the pretrained ResNet50 features. On the other hand, TLVQM \cite{tlvqm} and VIDEVAL \cite{videval} provide video level feature representations. We also experimented with video level features from numerous video clips using the features from the TLVQM \cite{tlvqm} model to evaluate it in a completely blind setting. However, the quality of the videos predicted using distance measure in Equation (\ref{distance}) from these features of the clips tends to perform poorly in terms of correlation with human scores. 
\begin{table}
\caption{SROCC performance comparison of VISION against other state-of-the-art VQA methods under linear evaluation protocol. Entry marked '-' denotes data is unavailable.}
\centering
\label{tab:lin_eval_vqa}
\begin{tabular}{|c|c|c|c|c|}
\hline
Data Type & KoNVid-1K & LIVE VQC & LIVE QCOMM\\
\hline
ResNet50 &0.670 & 0.653 & \textbf{0.749} \\
TLVQM & \textbf{0.755} & \textbf{0.758} & 0.727\\
VIDEVAL & 0.738 & 0.731 & 0.647\\
VCORNIA &0.475 &0.526 &0.462\\
HEKE &0.678 & 0.648 & 0.609\\
CSPT &0.702 & 0.623 & -\\
VISION &0.724 & 0.698 & 0.623\\
\hline 
\end{tabular}
\end{table}

\textbf{Feature Complementarity Analysis}. We study the complementarity among the features extracted from each stream by predicting the absolute error in quality prediction. For a given test video with human opinion score $Q$, let the error in quality estimation with features $(z_{f,d})$, and  $(z_{d,o})$ be given by $E_{f,d} = |Q - Q_{f,d}|$, and $E_{d,o} = |Q - Q_{d,o}|$, respectively. In Table \ref{tab:complementary}, we show consecutive frames from two videos from KoNVid-1K \cite{konvid} dataset mainly corrupted with illumination change and stabilization error respectively. We see that $E_{f,d}$ is higher than $E_{d,o}$ for the video mainly corrupted by illumination variation, while the trend is opposite when the distortion is due to motion only. We infer that when spatial distortion is predominant, using $(z_{f,d})$ is more useful due to the presence of frame level spatial information. While for the other video where temporal shakiness is predominant, $(z_{d,o})$ is more useful as optical flow information helps in understanding the distortion due to large motion. Thus, a combination of both sets of features for predicting quality will provide a better estimate for all types of distortion.

\begin{table}
\caption{Analysis of complementarity between frame-frame difference $(z_{f,d})$, and frame difference-optical flow-based features $(z_{d,o})$. Three consecutive frames from two videos in KoNVid-1K \cite{konvid} dataset are given in each row.}
\centering
\label{tab:complementary}
\begin{tabular}{|c|c|c|c|c|}
\hline
 frame 1 & frame 2 & frame 3 & $E_{f,d}$ & $E_{d,0}$\\
\hline
\includegraphics[width =0.2\columnwidth]{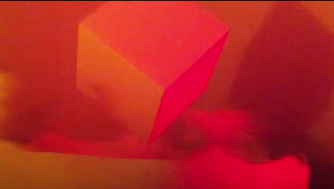} &\includegraphics[width =0.2\columnwidth]{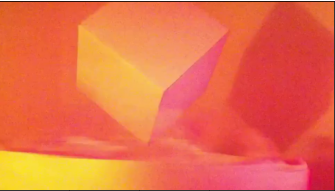} &\includegraphics[width =0.2\columnwidth]{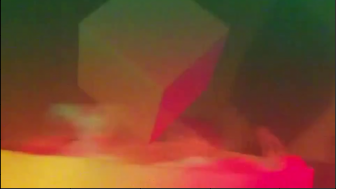} & 0.52 & 4.02\\
\hline
\includegraphics[width =0.2\columnwidth]{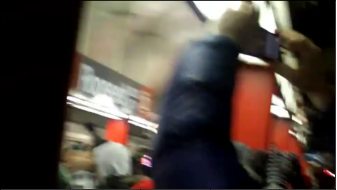} &\includegraphics[width =0.2\columnwidth]{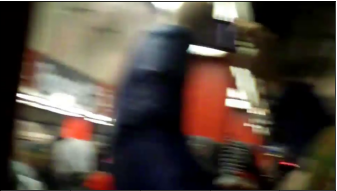} &\includegraphics[width =0.2\columnwidth]{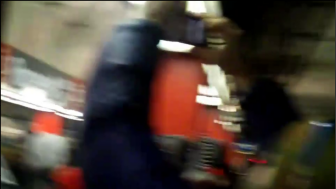} & 1.82 & 0.61\\
\hline 
\end{tabular}
\end{table}

\subsection{Runtime Analysis}
\begin{figure}
    \centering
    \includegraphics[scale=0.5]{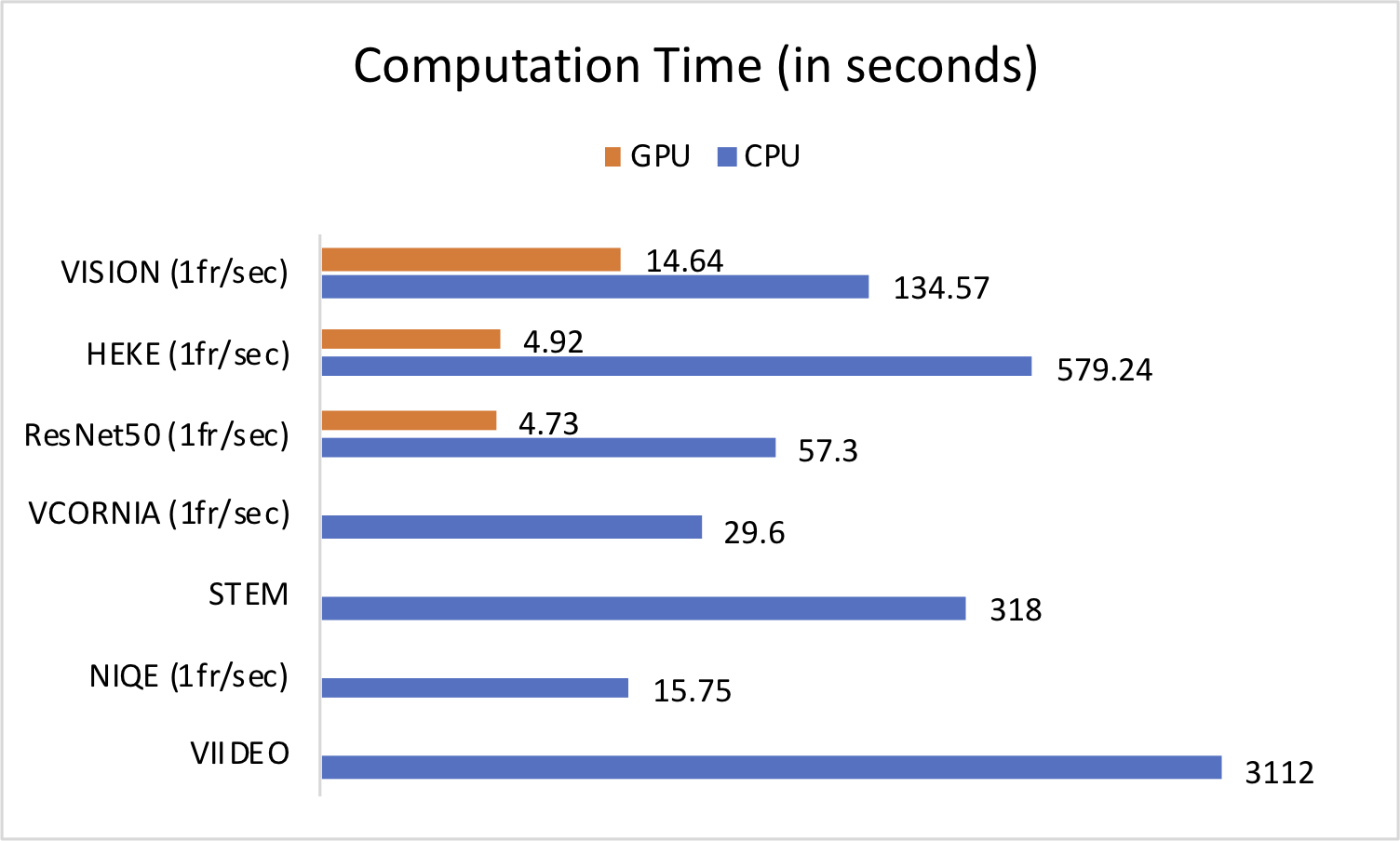}
    \caption{Runtime Comparison}
    \label{fig:computation}
\end{figure}

We present an analysis of the computational complexity of different methods through a runtime analysis. The runtime analysis is done on a Ubuntu 18.04.4 LTS system with a Intel® Core™ i7-8700 CPU @ 3.20GHz × 12. We also show the computational time for CNN-based methods using a 11GB GeForce RTX 2080 Ti graphics card. Since learning with 1 frame per second is an important aspect in our algorithm, we show runtime for VISION, $HEKE^*$, $VCORNIA^*$, NIQE \cite{niqe}, and pretrained $ResNet50^*$ at 1 frame/second.  The average computational time to predict quality for 10 videos from the LIVE Qualcomm \cite{liveqcomm} dataset consisting of 450 frames at $1920 \times 1080$ resolution is measured. In Figure \ref{fig:computation}, we see that on a CPU device VISION takes a similar time as that of other benchmarking methods. On a GPU platform, the CNN based methods are $10-15$ times faster than that on CPU device. As the time taken to compute the optical flow is around 6 seconds, VISION's runtime is slightly higher than other CNN based methods on a GPU.

\section{Conclusion}
We designed a framework for completely blind NR VQA of user generated videos using a multiview contrastive setup. We learn a quality aware feature representation by leveraging the common quality information in multiple views of a video. A combination of quality aware features from frame and frame difference, as well as frame difference and optical flow is used in a blind fashion to predict video quality. We show that these features can be used to compute an effective distance measure between the test video and a corpus of pristine videos to predict video quality. VISION gives a stable and consistent performance across all UGC datasets compared to other completely blind algorithms.

\clearpage
\bibliographystyle{ACM-Reference-Format}
\bibliography{paper}

\clearpage
\appendix
\begin{figure*}
    \centering
    \includegraphics[width=\textwidth]{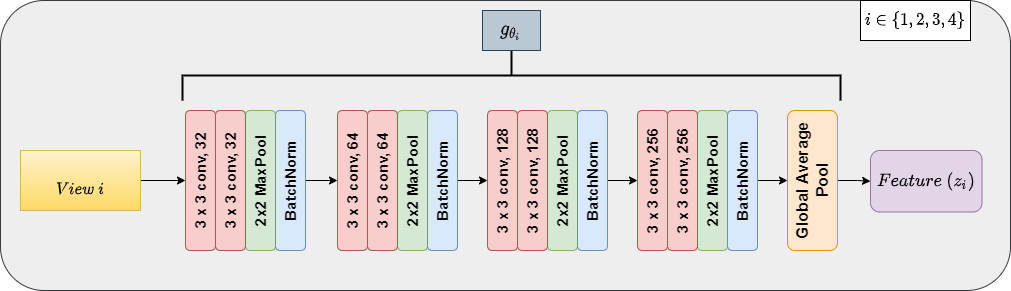}
    \caption{Architecture of our encoders $g_{\theta_1}$, $g_{\theta_2}$, $g_{\theta_3}$, and $g_{\theta_4}$.}
    \label{fig:encoder}
\end{figure*}

\section{Feature Encoder Architecture}

As mentioned in Section 3.3 in the main paper, we deploy a pair of convolutional neural networks (CNN) for each stream of our two-stream feature extraction module. To learn feature embeddings from frames, and frame-differences, we deploy $g_{\theta_1}$, and $g_{\theta_2}$. Similarly, to learn representations from frame-differences, and optical flow, we deploy $g_{\theta_3}$, and $g_{\theta_4}$. All the four CNN based encoders have the same architecture as shown in Figure \ref{fig:encoder}. Each encoder includes four convolutional blocks consisting of two convolutional layer followed by maxpooling layer, and a batch normalization layer. To extract feature representations, a global average pooling layer is applied at the output of the last convolutional block. The convolutional layers' kernels and biases are initialzed with random normal initializer. We used a rectified linear unit (ReLU) activation after each convolutional layer. 

\section{Impact of Frame Sampling}

In this section we study the impact of varying the frame sampling rate. VISION estimates video quality by averaging the predicted frame level quality at 1 frame/second instead of averaging over the whole duration of the video. We conduct a study to check the variation in performance of VISION on KoNVid-1K \cite{konvid} dataset due to the different frame sampling rate. We also provide the average computational time required to estimate the quality of videos. This study is done on an Ubuntu 18.04.4 LTS system with a Intel® Core™ i7-8700 CPU @ 3.20GHz × 12 with 11GB GeForce RTX 2080 Ti graphics card. In Table \ref{tab:frame_sampling}, we see that the performance is nearly similar as the sampling rate increases from 1 frame per second. There is a considerable dip in performance when the sampling rate is reduced below 1 frame per second. Thus, predicting the video quality by average pooling the estimated frame level quality at 1 frame per second gives similar performance while the average computational time is 9 times faster on a GPU machine.

\begin{table}[h]
\caption{Performance of VISION with different frame sampling rate on KoNVid-1K \cite{konvid} database.}
\centering
\label{tab:frame_sampling}
\begin{tabular}{|c|c|c|}
\hline
Sampling Rate & SROCC & Computational Time\\
& & (in seconds) \\
\hline
1 frame/4 sec & 0.5336 & 1.56\\
1 frame/2 sec & 0.578 & 1.96\\
1 frame/sec & 0.598 & 2.86\\
2 frame/ sec & 0.599 & 4.81\\
All frames & 0.601 & 25.85\\
\hline
\end{tabular}
\end{table}

\end{document}